\documentclass[conference]{IEEEtran}
\def\baselinestretch{1.05}
\usepackage{multicol}

\usepackage{graphicx}
\graphicspath{{eps/}}

\ifCLASSINFOpdf
\else
\fi
\usepackage{graphicx, amssymb, amsmath, subfigure}
\IEEEoverridecommandlockouts

\begin{document}

%
\title{Location Cheating: A Security Challenge to Location-based Social Network
Services\thanks{1. We have obtained consent from foursquare to
reveal the findings described in this report.}\thanks{2. This work
has been accepted by the 31st International Conference on
Distributed Computing Systems (ICDCS'11)}\thanks{3. The authors
contributed equally to this report and are listed in alphabetical
order.}}


\author{\IEEEauthorblockN{Wenbo He}
\IEEEauthorblockA{Electrical Engineering Department\\
University of Nebraska-Lincoln\\
Email: wenbohe@engr.unl.edu}
\and
\IEEEauthorblockN{Xue Liu}
\IEEEauthorblockA{School of Computer Science\\
McGill University \\
Email: xueliu@cs.mcgill.ca}
\and \IEEEauthorblockN{Mai Ren}
\IEEEauthorblockA{Computer Science and Engineering\\
University of Nebraska-Lincoln\\
Email: mren@cse.unl.edu}}

\maketitle

\begin{abstract}

Location-based mobile social network services such as foursquare and
Gowalla have grown exponentially over the past several years. These
location-based services utilize the geographical position to enrich
user experiences in a variety of contexts, including location-based
searching and location-based mobile advertising. To attract more
users, the location-based mobile social network services provide
real-world rewards to the user, when a user checks in at a certain
venue or location. This gives incentives for users to cheat on their
locations. In this report, we investigate the threat of location
cheating attacks, find the root cause of the vulnerability, and
outline the possible defending mechanisms. We use foursquare as an
example to introduce a novel location cheating attack, which can
easily pass the current location verification mechanism (e.g.,
\emph{cheater code} of foursquare). We also crawl the foursquare
website. By analyzing the crawled data, we show that automated large
scale cheating is possible. Through this work, we aim to call
attention to location cheating in mobile social network services and
provide insights into the defending mechanisms.

\end{abstract}

\section{Introduction}

A recent surge of location-based services (LBS) led by
foursquare\cite{foursquare:website}, Gowalla\cite{gowalla:website},
GyPSii\cite{gypsii:website}, Loopt\cite{loopt:website}, and
Brightkite\cite{brightkite:website} has attracted a great deal of
attention. Take foursquare as an example, it has become one of the
top recommended applications for all smartphone platforms. Till
August 2010, foursquare had attracted 1.89 million users since its
launch in March 2009, and it draws in more than 10,000 new members
daily. Meanwhile, hundreds of other similar services have been set
up to follow this growing trend.

To encourage the use of location-based social network services, the
service providers offer virtual or real-world rewards to a user if
he or she checks in at a certain \emph{venue} (i.e., places like
coffee shops, restaurants, shopping malls). Foursquare provides
real-world rewards (i.e., a free cup of coffee from Starbucks),
which gives users incentives to cheat on their location information
so that they can check in at a venue far away from where they really
are.

In this report, we use foursquare as an example to investigate the
vulnerability in location-based social network services. The goal is
to raise awareness of location cheating and suggest possible
solutions to drive the success of the business model among service
providers, registered venues, and users.

We first introduce a novel and practical attack on location
cheating, where a user may claim he or she is at a certain location
which is thousands of miles away from his/her actual location,
thereby deceiving the service provider on location information.
Though foursquare has adopted the \emph{cheater code} to stop
location cheating, we show that an attack can easily pass the
\emph{cheater code}. This benefits the attackers in the real world
and can be more severe when combined with the analysis on venue (or
location) profiles. In order to study foursquare's vulnerability to
location cheating, we also crawled the foursquare website and used
the crawling results to find suspicious cheaters on foursquare.

We found that the root cause of the vulnerability to location
cheating is the lack of proper location verification mechanisms. If
a user explores the open source operating systems for smart phones
(e.g., Android) to modify global-positioning-system-(GPS)-related
application programming interfaces (APIs), the user is able to cheat
on his/her location using falsified GPS information. Even if
defending mechanisms like \emph{cheater code} are deployed, the
loosely regulated anticheating rules still leave space for location
cheaters.

Our investigation suggests that defense against location cheating
requires improvement to location verification ability. We outline
the possible solutions to defense against location cheating. We
suggest service providers take the following measures to prevent
location cheating: (1) explore effective location verification
technologies, and (2) limit profile crawling and analysis to
mitigate the threat of location cheating. We believe that this
investigation on location cheating will have a great impact on
mobile social network services, and it will be an active research
topic with strong practical value.

The rest of this technical report is organized as follows. In
Section II, we briefly describe the background of LBS and its
associated business model, cheating scenarios and \emph{cheater
code}. In Section III, we introduce a basic location cheating
attack, demonstrate how to automate the cheating, and optimize the
benefits to the attacker through crawling and profile analysis. In
Section IV, we show the results from our experiments on location
cheating and examine the seriousness of current cheating threats. In
Section V, we discuss possible solutions to prevent location
cheating. In Section VI, we provide our conclusions and discuss
future work.

\section{Background}

In this section, we provide background and describe current
practices used by location-based mobile social network services.

\subsection{Business Model}

Location-based social networking services allow users to share their
location-related information. Users can add comments about a
restaurant, find out what's happening, let their friends know where
they are, and meet friends nearby for a cup of coffee. To report the
geolocation to a service provider (e.g.,
foursquare\cite{foursquare:website}), a user needs to ``check in''
to the location/venue where the user is located. The service
provider may broadcast the user's location information to his/her
friends or even the public. The check-in is done by hand, which
means a user is able to determine when he/she wants to check in,
thereby controlling their location privacy. Services like this are
not new, but they all have lacked incentives for people to use them,
until foursquare introduced a new business model.

Foursquare uses a progressive reward mechanism to provide four types
of reward incentives to its users. From the easiest to the hardest
to get, they are: points, badges, mayorships, and real-world
rewards. The first three are virtual rewards: (1) \emph{points} are
provided for all valid check-ins (e.g., the first time checking into
a venue, checking into the same venue multiple times); (2)
\emph{badges} are awarded for specific achievements, such as ``30
check-ins in a month'' or ``checked into 10 different venues''; and
(3) \emph{mayorship} of a venue is granted to the user who checked
into that venue the most days in the past 60 days. Only the number
of days with check-ins to this venue is counted, without
consideration of how many check-ins occurred per day or the total
number of check-ins. Unlike points and badges which depend solely on
a user's activities, the title of ``Mayor'' is given on a
competition basis. There is only one mayor at each venue. This
created a vulnerability in that if an attacker got the mayorship of
this venue and kept checking into it every day, no other users could
get the mayorship from the attacker.

The real-world rewards, like a free cup of coffee, were provided by
businesses (e.g., restaurants or bars) who set up a partnership with
foursquare. We crawled the information for all venues (discussed in
more details later) and found that more than 90\% percent of the
rewards were only for mayors. This setup provides benefits to both
foursquare and its partner businesses. On the other hand, the user's
desire for discounts and the competition for the mayorship will
likely bring more users (customers) to the partner businesses,
increasing their profits. While the business model benefits multiple
parties in the game, it makes foursquare a lucrative target to
attack by location cheating.

\subsection{Possible Location Cheating Scenarios}

In the context of location-based social network services, a user may
cheat on his/her location for various reasons. A user may want to
get rewards from venues or impress others by claiming a false
location. A business owner may use location cheating to check into a
competing business, and badmouth that business by leaving bad
comments.

Similar to most location-based social network services, foursquare
initially relied on users' self-regulation to maintain the
authenticity of the check-ins. Hence, the check-ins to any place a
user can find in the foursquare \emph{client application} (either
using the suggested list of nearby venues, searching for a venue by
name, or browsing and locating the venue on the map) were valid.
Software is available on the market that can automatically check
people into their desired venues, e.g., ``Autosquare'' for Android.
The basic cheating method worked in the early days of foursquare. It
is rather simple but obviously does not work now after the
introduction of the location verification mechanism, which requires
location information to complete the check-in process. However, the
location cheaters can modify the location information and send a
false location to the server.

The objective of the attacks is to automatically check into as many
businesses as possible and as frequently as possible to maximize
benefits through location cheating. A more sophisticated attack is
automated cheating. To make automated cheating easier, the cheaters
may use venue profile analysis to identify victims, which can be the
venues who provide discounts or users who aim to get mayorship in
specific venues. Hence, an attacker is able to select the venues
where the ``Mayor'' title is less competitive and the rewards are
more desirable or use a minimum number of check-ins to prevent
another user from getting a mayorship.

\subsection{Cheater Code}

Foursquare has adopted the \emph{cheater code} to defend against the
location cheating attacks. One of its functions is to verify the
location of a device by using the GPS function of that device. If a
user claims that he/she is currently in a location far away from the
location reported by the GPS of his/her phone, the check-in will be
considered invalid and won't yield any rewards.

Apart from utilizing GPS for location verification, the
\emph{cheater code} also incorporates multiple rules which run on
foursquare servers to determine if a user cheats on location. The
details of the \emph{cheater code} are concealed from users. But we
managed to detect a few rules, through experiments, that are
important to maneuvering location cheating to pass the scrutiny of
the \emph{cheater code}. A few of the criteria used in determining
location cheating in the \emph{cheater code} are listed as follows.

\textbf{Frequent check-ins}: We found a user cannot check into the
same venue again within one hour. This rule prevents a user from
checking in frequently to get as many points as possible and keep
his/her name on top of the recent check-in list, making it more
likely for people to contact the user for comments about the venue.

\textbf{Super human speed}: If a user continuously checks into
locations that are located far away from each other, foursquare will
indicate that the user is moving at ``super human speed'' and refuse
to give any reward for his/her check-ins. This rule limits location
cheating by a single user to a small geographic area.

\textbf{Rapid-fire check-ins}: If a user checks into multiple venues
located within a 180 meter by 180 meter square area (which is well
within a short walking distance such as in a mall) within a 1 minute
interval, foursquare issues a warning about ``rapid-fire check-ins''
on the fourth check-in. This rule stops a user from checking into
multiple venues in a small area and within a short time period.

These rules essentially limit the number of check-ins a user can
perform per day, thus reducing the potential for automated cheating.
Clearly identifying these rules helps attackers to design the best
way to work around them.

\section{Location Cheating Attack}

In this section, we outline three levels of attack: cheating via
GPS, automated cheating, and use of venue profile analysis to assist
cheating. They severely interrupt the operation of LBSs when
combined together. We first introduce four location cheating methods
which can pass the validation from foursquare and other similar
location-based social network services at least once. After that, we
crawl data from foursquare's website and evade foursquare's
\emph{cheater code} to automate the cheating process. Finally, by
analyzing the crawled data, we focus on a cheating attack on
high-valued targets such as those who provide real-world rewards.

\subsection{Location Cheating Against GPS Verification}

\begin{figure}
\centering
\includegraphics[width=80mm]{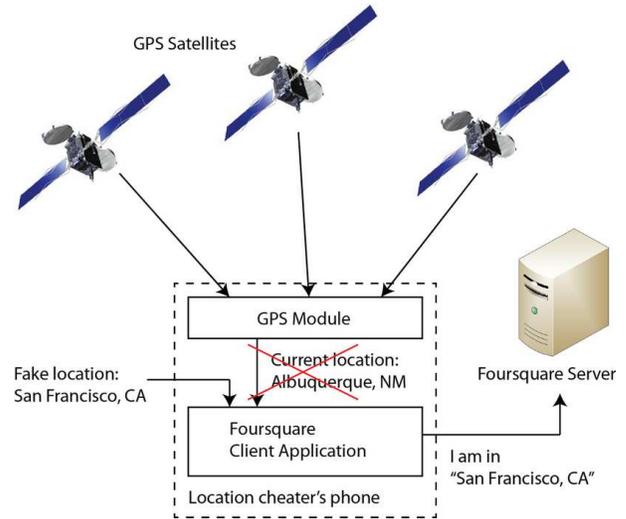}
\caption{Illustration of location cheating}
\label{fig:1}
\end{figure}

Location-based services like foursquare use their \emph{client
applications} installed on their users' smartphones to get GPS
location readings. Since this happens completely on the client side,
it is relatively easy to hack. We analyzed foursquare's \emph{client
application} source code and confirmed that it gets the GPS location
data from the phone's GPS-related APIs. Figure~\ref{fig:1} shows the
concept of such location cheating. Normally, the GPS module in a
mobile phone will return the current location information to the LBS
application, but an attacker blocks this and feeds fake location
information to the LBS application so that it makes its server
believe that this phone is really in the fake location. The cheating
check-in will then be approved.

There are several ways for an attacker to pass the GPS verification
by providing foursquare's \emph{client application} with fake GPS
coordinates:

\begin{enumerate}
\item \emph{Via GPS APIs}:

This method modifies the GPS-related APIs in a smartphone's
operating system to let it return fake GPS data. This is easy
because of the prevalence of open source smartphone operating
systems like Android. These APIs can be modified to get GPS
locations from sources other than the phone's GPS module, for
example, from a server that returns fake GPS coordinates or simply
from a local file. This method is limited to open source operating
systems; but since LBSs like foursquare provide their \emph{client
applications} on major smartphone platforms (Android, iPhone,
Blackberry), this is a universal cheating method. A hack into
Android is representative to cover cross-platformed LBSs.

\item \emph{Via GPS module}:

Directly hacking into a smartphone's GPS module is another way of
cheating on location. There are two ways to do this: one via
hardware and the other is via software. The former modifies the
physical GPS hardware inside the phone, making it capable of faking
data, so that the cheating is transparent to the mobile phone's
operating system. The latter simulates a GPS device. For example, an
attacker can write a program on a computer that simulates the
behavior of a Bluetooth GPS receiver and let the phone connect to
this simulated Bluetooth GPS receiver, enabling the simulated GPS to
return fake coordinates. In fact, there are already a number of such
tools on the market (e.g., Skylab GPS
Simulator~\cite{skylab:website}, Zyl Soft~\cite{zyl:website}, GPS
Generator Pro~\cite{gpsGeneratorPro:website}), that were originally
developed to help debug GPS-related software or gadgets.

\item \emph{Via server provided APIs}:

Foursquare provides a set of application APIs that allow developers
to create new applications for them, like an application for
uploading geotagged photos. These APIs can be employed by a location
cheater to check into a place. The drawback is that not all LBS
service providers provide such public server APIs. But this method
is more convenient to issue a large-scale cheating attack.

\item \emph{Via device emulator}:

Smartphone manufacturers (like Apple, Google, and Microsoft) provide
device emulators to developers for easier debugging and testing. A
device emulator is a full featured virtual machine of that device.
One of the basic features of these device emulators is that they are
configurable, including their simulated GPS module. Take the Android
device emulator, for example. We can send it a specific command to
set a location to the simulated GPS module. The GPS module of this
emulator will return the coordinates we set to whichever application
that needs GPS info. We conducted our experiments with this method,
because this one is the easiest and most reliable when compared to
the first two methods. Almost all potential attackers with a basic
knowledge of mobile developing can master this method with no
difficulty.
\end{enumerate}

We chose an Android emulator to conduct our experiments. 
We've registered a user on foursquare for testing purposes and
conducted all of our experiments in Albuquerque, New Mexico, and
Lincoln, Nebraska. Our goal was to check into venues outside of the
two states, so that we knew the cheating method was working. We used
the tool ``Dalvik Debug Monitor'', which is part of Android SDK to
connect to the emulator and set GPS coordinates in it. We found the
coordinates of the target venues by looking up Google Earth, which
shows the exact coordinates of where the mouse is pointing on its
map.

The entire cheating process can be described as: hack the emulator;
install and run foursquare application; find the coordinates of the
target venue in Google Earth; use ``Dalvik Debug Monitor'' to set
the coordinates in the emulator; find the target venue in the list
of nearby venues in the foursquare application; and check into the
target venue.

The results of our experiments showed that the check-ins to distant
venues were all accepted, and we received rewards successfully. We
got points for each of the check-ins, and we got badges like a
normal user as well, i.e., after checking into ten different venues,
we got the badge, ``Adventurer: You've checked into ten different
venues!''. We also tried to get a mayorship, we chose the venue
``Fisherman's Wharf Sign'' in San Francisco, which is a well-known
tourist spot, as the target venue; and we kept checking into it once
a day for four consequence days. After nine days, we found our test
user became the mayor of the venue. This experiment shows that the
\emph{device emulator method} works and can receive rewards.

\subsection{Crawling Data From Foursquare Website}

Getting the big picture of foursquare users and venues greatly helps
in location cheating attacks, although the crawling itself is not an
attack. There are two types of information that we crawled: users'
profiles and venues' profiles. In this section, we describe the
crawling procedure, in which we only accessed foursquare's public
webpages. Wondracek, Holz, Kirda and Kruegel
\cite{wondracek2010practical} introduced a similar crawling and
attacking approach. We will also use the crawling results to show
evidence of location cheating attacks on foursquare and identify the
suspicious location cheaters in the next section.

To increase performance, we developed a multi-thread crawler to
download and process a large amount of webpages (over 7 million).
This architecture has been proven to be highly effective, for
example, Cho, J. and Garcia-Molina used a parallel crawler to
increase performance \cite{cho2002parallel}, and Chau, Pandit, Wang,
Faloutsos focused on crawling social networks with parallel crawling
\cite{chau2007parallel}.

We wrote the crawler in C\# and used MySQL as the database. We ran
the crawler on three Windows PCs at the same time, each with a 2.0
GHz Intel Core 2 Duo processor and 1GB RAM. The fourth computer with
the same hardware specification, but running Ubuntu 8.10 server
operating system and functions served as a database server. In our
design, we set 14 to 16 threads on each of the three crawling
machines to crawl 100,000 users per hour for user profile crawling,
and set 5 to 6 threads on each machine to crawl around 50,000 venues
per hour for venue profile crawling.

In total, we crawled more than 1.89 million users and 5.6 million
venues, which agrees with foursquare's reported number of users.
This means we can update all user profiles in less than two days or
update all venue profiles in about five days. The crawling
performance is an important design concern, because by repeatedly
crawling data and comparing the differences between each set of
crawling results, we can further investigate the behaviors of its
users and extract more information. For example, the venue's recent
visitor list does not have a time stamp to indicate when a user
visited this venue; but if we crawl the venues daily, then we will
be able to determine how frequently a user checks into a venue. We
can further analyze the user behavior to show if the user is
suspicious for location cheating.

Each user on foursquare has a profile that contains personal
information. A user's profile provides information such as name,
current location, check-in numbers, reward information, and list of
friends. We cannot access the mayorships and check-in history
directly (i.e., they are hidden from the public), since these two
types of information may expose his/her location privacy. However,
we can infer a user's mayorship information and partial check-in
activities from venue profiles, which contain lists of recent
visitors, and links to mayors. In addition, a venue profile also
provides its name, address, location, number of users who checked
in, unique visitors, and tips.

To crawl these profiles, we need to know the URLs of these profile
pages. We discovered that foursquare uses continuously numbered IDs
to identify their users and venues. By changing the ID in the URL,
we can crawl almost all of the user and venue profiles. We believe
this is a serious security weakness and should be patched soon.

Two types of URLs can be used to access user profiles. The first one
is with an internal user ID in URL, like
``http://foursquare.com/user/-1852791''. To access another user with
ID 23456, we just replace the ``1852791'' in the URL with ``23456'',
and we can visit the public profile page with the new URL. We
believe that all the users are accessible just by increasing or
decreasing the user ID in the URL. We implemented a web crawler to
do so, and we discovered around 1.89 million users in August 2010.
Another type of URL contains the username, like
``http://foursquare.com/user/test'', where ``test'' is the username
of a user. Not all users have a username-based URL as their profile
page. Out of 1.89 million users, only 26.1\% have a username, so we
used the URL with ID in our crawling tool. For venue profiles,
foursquare only uses numbered IDs in the URL of the profile pages,
like ``http://foursquare.com/venue/1235677''.

\begin{figure}
\centering
\includegraphics[width=80mm]{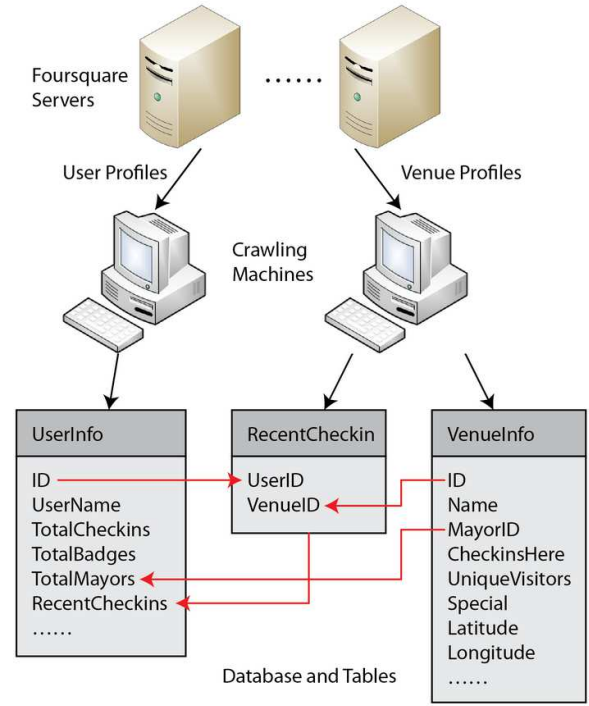}
\caption{A crawling architecture and the database to store crawled
information from foursquare} \label{fig:4}
\end{figure}

After we had the URL of a profile page, we sent HTTP Get to this URL
and got the HTML source code from the server's response. To extract
data from the HTML source code, we let the crawler perform a set of
regular expression matches. After extracting the data, we stored
user and venue information in a database.  Figure~\ref{fig:4} shows
the structure of the database; the arrows indicate the relationships
between the tables. We stored user and venue profiles in tables
\emph{UserInfo} and \emph{VenueInfo} respectively; and we also
created a table called \emph{RecentCheckin} to record the relations
between venues and users. We put each venue's recent visitors in
this table; and by counting the number of records for a user, we
recorded the number of recent check-ins of this user and stored it
in \emph{RecentCheckins} of \emph{UserInfo}. Similarly, by analyzing
the \emph{MayorID} of each venue, we calculated how many mayorships
each user had and put the result in \emph{TotalMayors} of
\emph{UserInfo}.

\subsection{Automated Cheating}

To achieve significant benefits from location cheating, attackers
need to be able to control a large number of users and make them
check in automatically. This requires the location cheaters to (1)
find location coordinates of victim venues by computer program, and
(2) automatically select a list of venues to check into pass the
\emph{cheater code}. We met the first requirement by crawling, and
we could easily use SQL commands to get the location coordinates of
the selected venues from the database.

\begin{figure}[h]
\centering
\includegraphics[width=80mm]{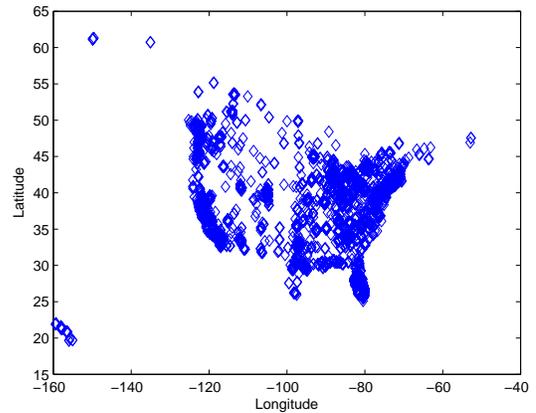}
\caption{Locations of Starbucks branches crawled from foursquare
website} \label{fig:2}
\end{figure}

Figure~\ref{fig:2} shows the coordinates of all Starbucks branches
in the US, where x axes and y axes are real coordinates. The
location coordinates form the shape of United States territory,
because Starbucks' branches are distributed all over the US. We draw
this map by SQL command:

\emph{SELECT Longitude, Latitude FROM VenueInfo WHERE Name LIKE ``\%Starbucks\%''}.

Second, to pass the anticheating verification, the key is to avoid
triggering any of the rules in the \emph{cheater code}, since it
detects cheater on a per user basis, we focus on the strategy of a
single user. An attacker needs to organize coordinates from the
first step into a schedule, which states the sequence of venues to
check into and the time interval a check-in has to wait after the
previous check-in; and the schedule must follow all rules from the
\emph{cheater code}. The attacker could create a tool to do this
automatically.

To determine the sequence of venues in which to check in, an
attacker can create a virtual user, compute a virtual path to visit
the target venues using Google Map's APIs, and build the check-in
schedule along the virtual path. We also need to determine the time
interval $T$ between check-ins, which is determined by distance
between the check-ins in the schedule. Based on our experiments, we
can check into venues less than 1 mile apart with a 5-minute
interval without being detected as a cheater. So for distance $D$,
less than 1 mile, we should set $T$ to 5 minutes; if $D>1$ mile, we
let $T = D*5$ minutes.

In our proof of concepts experiment, we created a semiautomatic
location cheating tool. With the tool, an attacker can use any venue
as the starting point. The attacker can then set the next cheating
location by setting the moving direction and distance, for example,
``move 500 yards to the west'', the tool will search for the venue
that is the closest to the target location and then automatically
set the coordinate of the found venue to the emulator or generate a
list for fully automated check-in later. The tool also automatically
limits the process to avoid triggering any anticheating rules in the
\emph{cheater code} as we presented before.

\begin{figure}
\centering
\includegraphics[width=80mm]{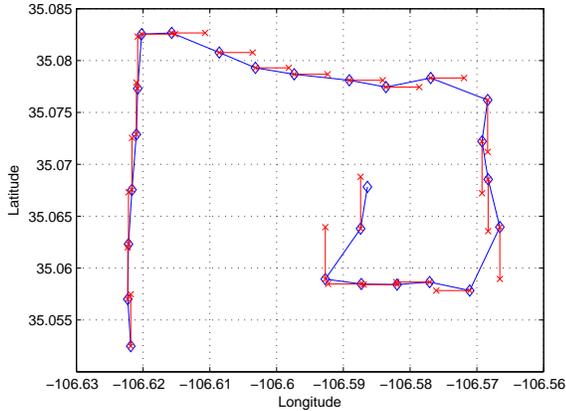}
\caption{An illustration of location cheating check-ins along a virtual path in the city.} \label{fig:3}
\end{figure}

Figure~\ref{fig:3} shows the path of a virtual tour, the diamond
points are the locations of venues the tool actually selected and
checked into, and the cross-points and lines to them show the
intended moving directions and target locations. The starting point
in this tour is at the lower left point of  Figure~\ref{fig:3}. We
started by moving north and then kept turning right. The desired
moving distance for each step was 0.005 degrees, either longitude or
latitude, equivalent to about 550 meters in latitude direction or
about 450 meters in longitude direction around this location. We set
the interval between check-ins to 5 minutes since the moving
distance is less than 1 mile. We continued checking into 25 venues
without being detected as a cheater, and we received reward points
and badges accordingly.

As we can see in Figure~\ref{fig:3}, for most of the time, the
actual venues we checked into are not very far from the desired
location, this is due to the high density of venues in the city. To
move across large distances, we should increase the moving distance
of each step, which will reduce the probability that we drift too
much from the desired direction, like the second to last move in
Figure~\ref{fig:3}.

\subsection{Cheating with Venue Profile Analysis}

Since the brute-force check-ins increase the chance that a cheater
is caught, the location cheaters may gain intelligence from the
venue analysis after the crawling. For example, an attacker may
select the victim venues that provide special offers to their mayors
and don't have a mayor yet (or are less competitive for mayorship)
as targets. It is relatively easy to become the mayor of these
venues. At the time this report is being prepared, around 1000
venues fall into this category.

Through profile analysis, we found a user in foursquare is the mayor
of 865 venues but with a total check-in number of just 1265. It is
interesting to observe that most of the 865 venues have no other
visitors during the past 60 days, so only one check-in is enough to
get the mayorship. We also discovered some special offers that do
not require mayorship which are much easier to get, it's hard to
find such information without crawling the venue profiles.

The attack can also target other users. For example, to stop a user
from getting any mayorship, the attacker will analyze the venue
profiles and find the venues that the victim user has been to or is
the mayor. Then the attacker will apply an automated cheating attack
on those venues in order to attack the mayorships of the victim.

\section{Evaluation of Location Cheating on Foursquare}

\begin{figure*}
\centering
\includegraphics[width=160mm]{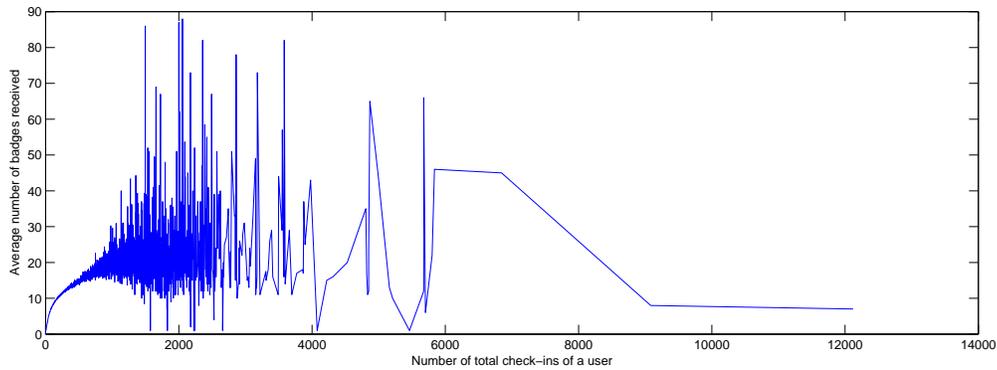}
\caption{Number of badges vs. number of check-ins: The average
number of badges granted to users who have a certain number of total
check-ins.} \label{fig:6}
\end{figure*}

We have demonstrated that location cheating attack on foursquare.
Next, we will show a big picture of location cheating through our
crawling and analysis. In this section, we examine the signs of
location cheating on foursquare. We found three identifying factors
that related to location cheating. They are: (1) above normal level
of activity, (2) below normal level of rewards, and (3) suspicious
check-in patterns.

\subsection{High Check-in Frequency in Recent Visitor List}

If a user checks in too frequently and at too many venues, it is
suspicious, because it is unlikely the user will visit too many
places in a short amount of time.

We crawled the record of 20 million check-ins, and each of them
represents a user visiting a venue once. That means, on average,
each user on foursquare has checked into at least ten venues, and a
venue has had at least four visitors. The actual number should be
higher since only recent check-ins were shown on the website and
were crawled.

\begin{figure}[h]
\centering
\includegraphics[width=80mm]{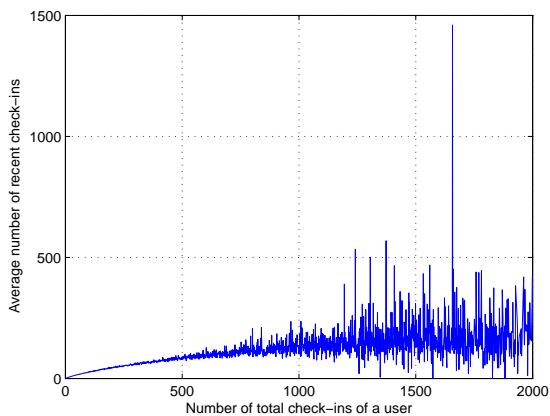}
\caption{Recent check-ins vs. total check-ins: the average recent
check-ins of the users who have a certain number of total
check-ins.} \label{fig:5}
\end{figure}

Figure~\ref{fig:5} shows a relationship between the number of total
check-ins and the recent check-ins. A recent check-in of a user
means that the user is in a venue's recent visitor list, but we
cannot directly know when this check-in happened. If this user is
the only visitor of this venue, then he will be staying in the
recent visitor list even if this check-in happened a year ago. In
fact, there are 1,291,125 venues that have only one check-in; and
2,014,305 venues have had only one visitor ever. Though it is not a
hard proof, the high ratio of {\it recent check-ins} to {\it total
check-ins} of a user indicates that it is likely a user plays tricks
to stay in the recent visits list, which is a sign of cheating.
Here, we only included users with 2000 or less total check-ins since
they cover 99.98\% users. We get the number of recent check-ins vs.
the number of total check-ins of each user, and then we compute the
average number of recent check-ins for users who have a given number
of total check-ins (see Figure~\ref{fig:5}). We can see that some
users with more than 1,000 check-ins have a unusually high
percentage of recent check-ins, which suggests that those users are
possibly cheaters, since it is not very likely for users to always
check into a large number of different venues in a short time
period.

From Figure~\ref{fig:5}, we can see that, on average, we get around
100 recent check-ins of a user, if the user did more than 500
check-ins total. There are 25,074 users that have a total check-in
number falling in between 500 and 2000. It's not difficult to
determine where they have been or are likely to go from this data.

\subsection{Low Reward Rate}

If a user has a large amount of check-ins but little rewards like
badges, the user may have been detected as a cheater by foursquare
so those check-ins were invalidated toward rewards, although they
still increase the total check-in numbers of those users under
foursquare's current policy. Figure~\ref{fig:6} shows the relation
between rewards (badges) and the number of check-ins. We first get
the number of badges vs. the number of total check-ins of each user,
and then we compute the average number of badges for users who have
a given number of total check-ins. As shown in Figure~\ref{fig:6},
for users with 1000 or less check-ins, the relation between the
number of check-ins and badges is stable. It illustrates that a user
will be likely to get more badges after doing more check-ins. It is
reasonable because the rewards are usually granted to those who have
checked in over a certain number of times to a venue. For the users
with a larger number of check-ins, we can see that the curve in
Figure~\ref{fig:6} oscillates dramatically. Actually, many users
with more than 1000 check-ins only have less than 10 badges. We
think the best explanation for this is that they are location
cheaters and got caught by foursquare and, thus, their check-ins
yielded no rewards. For almost all users with more than 9000
check-ins, the reward level is low. The average check-ins per day
for these users is over 16 times since the foursquare service was
launched in March 2009, which is a strong evidence that these users
are cheaters.

We notice that among 1.89 million users, 36.3\% have never checked
into any venues, 20.4\% have one to file check-ins, which means more
than half of the users have only checked in less than six times. On
the other hand, 0.2\% of the users have checked in at least 1,000
times; and 11 users have checked in at least 5,000 times. These 11
users who have made no less than 5,000 check-ins can be divided into
two distinct groups by the number of mayorships they have. The first
group has six users, each of whom is mayor of tens of venues, which
are all concentrated in a city area. The other five users in the
second group, including the one with over 12,000 check-ins, the
highest among all users, do not have any mayorships, and they
received much less badges than the first group. A further analysis
indicates that four of the five users in the second group appeared
in a recent visitor list of a venue, while the users in the first
group are all in the recent visitor lists of a large amount of
venues. This provides us with strong evidence that the users in the
second group are cheaters and were caught, so their check-ins are
invalidated.

\subsection{Suspicious Check-in Patterns}
Next, we will examine if the check-in pattern or history can tell if
a user is a location cheater through further analysis of the crawled
data.

\begin{figure}
\centering
\includegraphics[width=80mm]{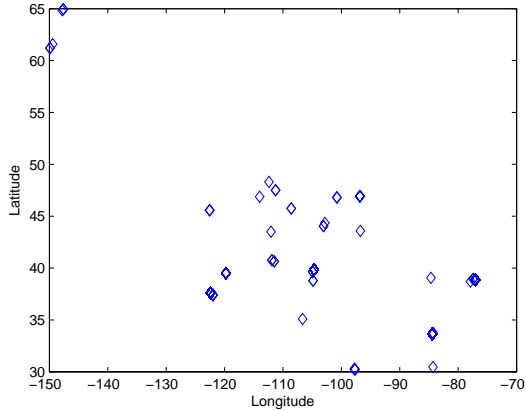}
\caption{Check-in locations of a suspected cheater} \label{fig:7}
\end{figure}

We analyze a user's check-in pattern based on the recent check-in
records. Figure~\ref{fig:7} shows the recent check-in locations of a
suspected cheater. We draw the venues to which a user has checked in
on a map, so that we have a general idea of the places the user has
``visited''. This user is in the recent visitor  lists of over 1000
venues. As we can see in  Figure~\ref{fig:7}, those venues are
scattered pretty far apart and spread over 30 different cities
throughout the United States, including Alaska, and Europe. Judging
from this user's ID (foursquare increments this ID as user
registers), we believe that the user has used foursquare for less
than one year. It illustrates that within a year, the user has
``visited'' at least 30 different cities, hence this user is
suspected of location cheating.

\begin{figure}
\centering
\includegraphics[width=80mm]{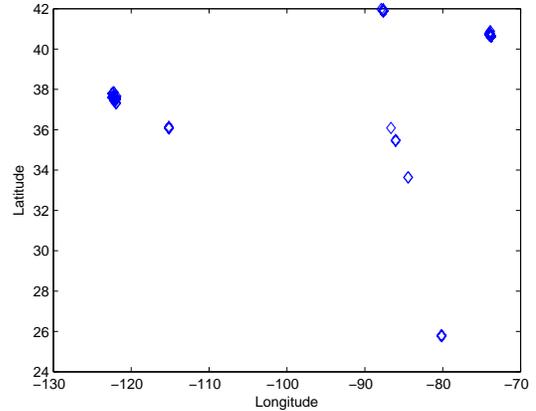}
\caption{Check-in locations of a ``normal'' user} \label{fig:8}
\end{figure}

Figure~\ref{fig:8} shows the recent check-in locations of a user
with a similar number of recent check-in records and similar ID (it
means the two users registered for the foursquare service at almost
the same time) as the user in Figure~\ref{fig:7}. But the venues
he/she visited are concentrated in three cities (places with darker
markers) and a few other places, where he may have visited on
vacation. After examining the users with more than 1,000 recent
check-in records, users with more than 2000 total check-ins, and
users with more than 100 mayorships, we believe that the check-in
pattern in Figure~\ref{fig:8} is normal.

In the future, we will focus on those cheaters that haven't been
detected by foursquare's current system. Foursquare implemented its
\emph{cheater code} anticheating system online around April 2010.
Since then, all detected cheating check-ins are still count in the
total number of check-ins, but do not receive any rewards. By the
time this work was conducted (August, 2010), all mayors passed the
scrutiny of the \emph{cheater code}. So any cheaters we found in
this group of users were new discoveries. There are 425,196 users
who have the mayor title, and there are 2,315,747 venues which have
mayors. On average, each user with a mayorship is the mayor of 5.45
venues. Those who are mayors of many venues are likely to be
cheaters.

\section{Possible Solutions Against Location Cheating}

The investigation presented in this work allows an attacker to
launch automated location cheating attacks against a large number of
victims, including service providers, business partners, and users.
The root cause of the vulnerability is the lack of effective
location verification mechanisms which can be deployed at a large
scale. However, it is possible to counter these attacks. In this
section, we list possible techniques to thwart the location
cheating; and we suggest that location security be enhanced by
limiting the access to user and venue profiles.

\subsection{Location Verification Techniques}

\noindent {\bf Distance bounding}: Distance bounding protocols
\cite{hancke2005rfid}, \cite{chiang2009secure},
\cite{sastry2003secure} that exploit the limitation on transmission
range or speed of a communication signal for location verification,
which does not rely on GPS inputs. This solution requires the
deployment of verifiers around the registered venues, hence it will
be expensive to deploy location verification based on distance
bounding.

\vspace{0.05in}

\noindent {\bf Address mapping}: Using address mapping to geolocate
IP addresses has been proposed in various applications, such as
\emph{Tracert Map} and \emph{Google Location Service}. Researchers
have adopted IP address mapping to locate mobile phones
\cite{balakrishnan2009s}. A challenge of applying IP address to
verify location is that mobile phones may access the Internet from
nonlocal IP addresses, and the IP addresses can be changed
dynamically.

\vspace{0.05in}

\noindent {\bf Venue side location verification}: The Wi-Fi routers
that provide the Wi-Fi hotspot services can work as location
verifiers. This technique provides an intrinsic distance bounding
since only devices that are physically within the radio
communication range of a Wi-Fi router can communicate with it.
According to previous literature~\cite{lehr2003wireless,
howard2006experimental}, the radio range of a Wi-Fi router is
generally no more than one hundred meters. This range level enable
use to identify cheaters that are miles away from the venue.
However, for the cheaters within the transmission range a Wi-Fi
router, this approach does not work. For example, a cheater sitting
inside a McDonald's can check-in to the Wendy's next door, which is
only 50 meters away. In this case, Wendy's owner can configure the
Wi-Fi router to limit the communicate within the restaurant via
hardware or firmware configuration tools (i.e., DD-WRT
\cite{ddwrt:website}). In this solution, a Wi-Fi router takes the
responsibility to measure if a check-in message was sent from a
device in a legal area by checking the communication delay between
the Wi-Fi router and the device. If so, the Wi-Fi router sends the
verification information to the corresponding LBS server.


In order to provide location verification service, the Wi-Fi routers
must be registered to the LBS server and establish trusted
communication with the server to block the impersonating attacks by
location cheaters.



When comparing the three solutions, \emph{Distance Bounding}
provides the most accurate location data, and it can be used
anywhere, but it is difficult to implement and has the highest cost.
\emph{Address Mapping} is the least accurate in terms of the
location data it provides, it can be used anywhere, and it has the
lowest cost and is the easiest to implement. \emph{Venue Side
Location Verification} has enough location accuracy, and it incurs
no extra hardware purchase or installation cost for the venues.
Owners of the venues can simply update the software on their
existing routers to make these routers capable of defeating location
cheaters.

\subsection{Mitigating Threat from Location Cheating}

As alluded above, with the assistance of profile analysis, an
attacker may optimize the location cheating strategies. To limit the
effect of potential location cheating attacks, we need to reduce the
information exposed to the public. Along this direction we can
employ the following techniques.

\vspace{0.05in}

\noindent {\bf Access control for crawling}: To prevent large-scale
profile analysis by attackers, a direct solution is to take counter
measures to stop or limit crawling. If a user must login to view the
publicly available profile pages, it will be easier to detect the
crawling users and block them. This can be combined with IP address
blocking, if the service provider can detect the crawler's IP
address. Even if the crawlers hide behind network address
translations (NATs), blocking their IP addresses will cause limited
collateral damage. Casado and Freedman \cite{casado2007peering} show
most NATs only have a few hosts behind them, and proxies generally
have much more. Crawling behind a public proxy cannot achieve enough
performance. Although tools like Tor \cite{dingledine2004tor} may
provide a high level of anonymity on the Internet, it also suffers
from limited performance for crawling purpose.

\vspace{0.05in}

\noindent {\bf Hiding information from profiles}: To reduce the
information leak, we hope that even if an attacker can successfully
crawl the website, the information that can be extracted from the
data is still limited. But if a subset of information in the
profiles is removed, the usability of the location-based social
networking service will be suffered. For example, if the recent
check-in list is removed from the venue's profile, users cannot
query the recent visitors to the venue for their comments about the
venue. Hence, removing the information from profiles is not a good
solution to prevent profile analysis. Rather, the service provider
may use the hash function to hide necessary information (such as
user IDs in the recent check-in list). Recently, the information
leak has been studied. Griffith and Jakobsson
\cite{griffith2005messin} use public records to infer individuals'
mothers' maiden names, and Heatherly et al.
\cite{heatherly2009preventing}, as well as Zheleva and Getoor
\cite{zheleva2009join}, show how public data provided by social
networks can be used to infer private information.

\section{Conclusions and Future Work}

In this technical report, we introduced a novel and practical
location cheating attack that enables an attacker to make the
location-based service providers believe that the attacker is in a
place far away from his/her real location. Through real world
experiments on foursquare, the leading location-based social
network, we demonstrate that our attacking approach works as
expected; and location cheating really threatens the development and
deployment of location-based mobile social network services. The
counter measures against location cheating in current systems are
not perfect.

We suggest several techniques for enhancing the security of location
information. In the future, we will investigate to find better
solutions to identify possible cheaters, especially those whom
haven't been found by the existing anticheating mechanisms. We also
would like to seek better solutions to the balance between the
usability and the security in order to make the location-based
mobile social networking service more attractive.

\IEEEtriggeratref{15}

\bibliographystyle{IEEEtran}
\bibliography{LocationCheating}

\end{document}